# Short range networks of wearables for safer mobility in smart cities

Kapil Sharma, Christian G. Claudel, *Member, IEEE*

*Abstract*— Ensuring safe and efficient mobility is a critical issue for smart city operators. Increasing safety not only reduces the likelihood of road injuries and fatalities, but also reduces traffic congestion and disruptions caused by accidents, increasing efficiency. While new vehicles are increasingly equipped with semi-automation, the added costs will initially limit the penetration rate of these systems. An inexpensive way to replace or augment these systems is to create networks of wearables (smart glasses, watches) that exchange positional and path data at a very fast rate between all users, identify collision risks and feedback collision resolution information to all users in an intuitive way through their smart glasses.

## I. INTRODUCTION

In the United States alone, more than 100 people die in traffic-related accidents per day, with an estimated 90% of traffic-related crashes caused by human error[1]. Motor vehicle collisions are the $6^{th}$ most common preventable cause of death, and the leading preventable cause of death in the age group 15-44. In most countries, vulnerable users (such as bicyclists or pedestrians) account for a disproportionately high number of traffic-related injuries and fatalities.

The emergence of autonomous vehicles will likely have a strong impact on vehicular safety, though these systems will likely require decades to be generalized. In the context of transportation in smart cities, safety is at the core of novel V2X (Vehicle to Vehicle or Infrastructure) communication systems[2], which are likely to become mandated in the coming years. Indeed, the primary purpose of these systems will be to broadcast a **Basic Safety Message** (BSM) to surrounding users over **Dedicated Short Range Communications** (DSRC), informing them on the future path of the vehicle (with position, velocity, acceleration and user input information). The purpose of these systems is to depart from the traditional paradigm of passive safety, in which vehicles and users are not equipped with systems helping them to avoid collisions. With these V2X systems, active collision warning and avoidance systems can be implemented on new vehicles, though the added cost of sensor and actuator systems will likely limit these features to high end vehicles.

In contrast, wearable devices such as smart glasses or smart watches will likely be massively adopted in future years[3]. Unlike smartphones, these devices offer us unprecedented opportunities in sensing, allowing the detection of hand gestures through the smart watch Inertial Measurement Units (IMUs), and head motion or visual scenes through smart glasses IMUs and cameras. This sensing data can be used to complement vehicle sensor data and create a more reliable collision detection and avoidance system.

Smart glasses can also be used to visualize information directly in the field of view of users, in a natural way, for example as boxes around users at risk of collision, or suggested paths in the form of trajectory holograms. Such devices could be used with any vehicle, and would replace expensive vehicle Head Up Display (HUD) systems for visualization, and simultaneously offer additional sensing capabilities such as tracking the orientation of the driver's head and generating images seen by the driver. This additional sensing would also come at no marginal cost for the drivers. On low-end or non-DSRC equipped vehicles, wearables could also be used as a standalone system, allowing an inexpensive retrofit of older or inexpensive vehicles with a semi-active traffic safety system (which could be viewed as a human-in-the-loop control system).

## II. PROTOTYPE SYSTEM

The Civil, Architectural and Environmental Engineering department of the University of Texas Austin is currently investigating a prototype traffic collision detection system based on a network of Google Glasses, Android smartphones and Android smartwatches exchanging and processing positional and attitude data through a cloud-based service. In this system, the positional data is generated by a fusion of GPS and IMU data, or by a high accuracy GPS system (such as D-GPS or RTK-GPS). This data is currently exchanged over short-range Bluetooth communications between users, which can then visualize people around them (and in particular people that are at risk of collision with them) using the smart glasses. The equipment used to develop this system is shown in Figure 1.

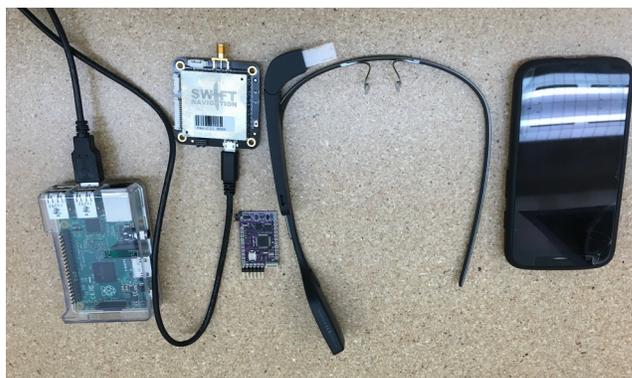

*Figure 1. Prototype equipment: Raspberry Pi embedded computer (left) connected to a SWIFT RTK-GPS (center, left), Google Glass, Samsung Android phone.*

*Research supported by the CAEE Department of UT Austin.

Kapil Sharma is with the Electrical and Computer Engineering department of the University of Texas, Austin (e-mail: kapil@utexas.edu).

Christian G. Claudel, is with the Civil, Architectural and Environmental Engineering department of the University of Texas, Austin (corresponding author, e-mail: christian.claudel@utexas.edu).

The attitude of the user heads can be inferred through a fusion of accelerometer, gyroscope, magnetometer and camera data generated by the smart glasses. The other wearables (for example smartwatches) can be used for gesture recognition, which helps in both predicting the users reachable sets and obtaining context information.

The anticipated system implementation is illustrated in Figure 2. This implementation relies on a cloud-based service to assess collision risk in real time from data preprocessed by all users. If a collision is detected, a resolution advisory can be computed and sent to each user for action, similarly to the aircraft TCAS (Traffic Collision Avoidance System) system [4]. The advisory can be visualized on each user glasses, for example in the form of a suggested tridimensional trajectory hologram, or augmented reality cues preventing some action (for example turning in a particular direction or accelerating).

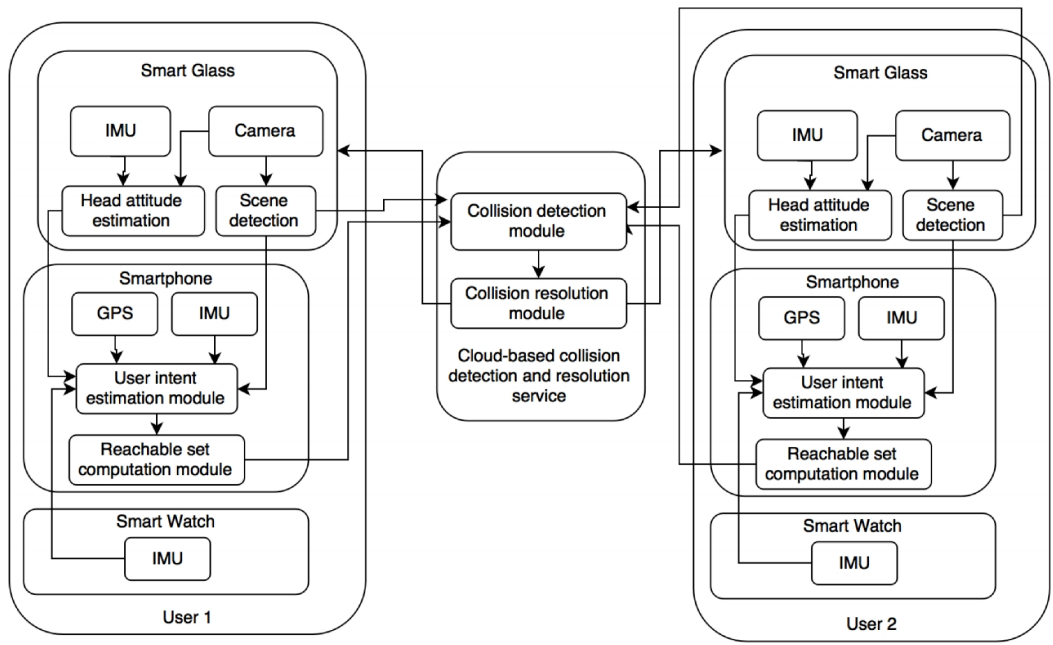

*Figure 2. System overview, on a two-user example with centralized collision detection and resolution. In this system, the measurement data generated by the wearables (smart glass, smart watch and smartphone) is first processed locally to estimate the positions and attitudes of the head and limbs, and perform basic context recognition[5], such as scene detection from video data, or transportation mode recognition (bicycle, pedestrian, motorcycle, car) using inertial data. This data is then sent to a user intent estimation module, which uses supervised learning in conjunction with real time data to estimate the future intents of the user. The set of intents is used to compute the reachable sets of each users, which are sent to a centralized collision detection module. If a collision is detected, a collision resolution is computed and is sent to users glasses for display.*

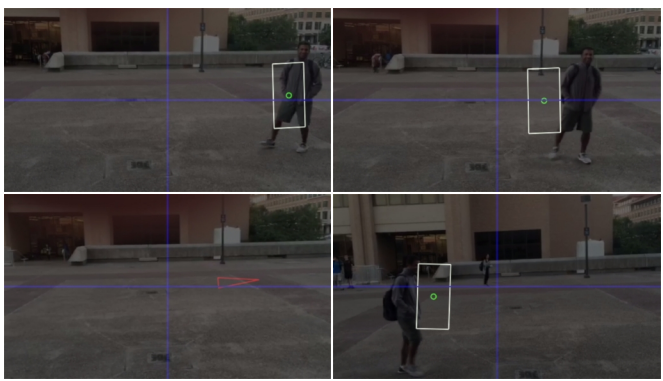

*Figure 3. Illustration of the positioning system developed at UT Austin. These screen captures show the view of a user equipped with a Google Glass, observing another user equipped with a smartphone streaming his positional data (obtained through GPS) to the smart glass over Bluetooth. Future versions of this system will fuse camera data generated by the glass to improve positional accuracy.*

The screenshots in Figure 3 illustrate the prototype system. These screenshots have been obtained by overlaying the position estimate generated by the Google Glass on video data of the scene. The corresponding video can be seen at: https://youtu.be/IxmklzzS5Ic

### III. RESEARCH CHALLENGES

#### A. Sensing

Positioning users accurately is critical for efficient active safety systems. In open environments, GPS data is usually accurate to a few meters, and random errors can be largely canceled by fusing inertial data to GPS measurements, though systematic errors persist, as can be seen from Figure 1. The accuracy of GPSs significantly degrades in urban environments however, requiring other types of absolute or relative positioning systems.

One of the greatest advantages of wearables is their capability of monitoring users much more accurately than cellphones (particularly if multiple wearables are used). These additional pieces of information are critical to traffic safety applications:

- Scene detection using cameras-equipped smart glasses allows the system to check if the user is looking in the right direction (for example if the driver of a vehicle can see another vehicle passing them in the mirrors or not), has seen a possible collision conflict or not (for example if a passing vehicle does not appear in the mirrors), or check if a possible collision conflict is hidden by an obstacle.

- Gesture detection using a combination of cameras and IMUs allow the system to identify the gestures of users, which can in turn be used to predict the path of users over some short time horizon[6]. Accurate path prediction would give more time for users to react, and reduce the likelihood of an accident actually happening, while limiting false-positive detection occurrences.

Detecting such features is however challenging. The computational power of smart glasses and smart watches is currently highly restricted, and extremely efficient learning and image processing algorithms will be required to perform the desired sensing operations in real-time on future systems.

### B. Networking

Several communication systems can be used for safety applications, including DSRC, Bluetooth, WiFi and low-latency cellular communication systems. DSRC and low-latency cellular systems are unfortunately not yet mainstream, though all vehicles will likely be equipped with DSRC within a few years. WiFi or cellular systems can be used to relay positioning or path information when conflicting users are very far away, while Bluetooth (which has a much lower latency) could take over when the collision risk is imminent.

The bandwidth required by such a system is expected to be low, since only positional, input and path information is expected to be exchanged between users through broadcast. However, low-latency is critical, with collision events taking place in split seconds. DSRC will have some reserved bandwidth for low-latency communication (for safety purposes), which can be leveraged if such systems become mainstream on wearables or smartphones.

### C. User path forecast

Forecasting the paths of all users accurately is of critical importance for a collision detection and resolution system. Indeed, relying on the current state of the system to forecast the reachable sets of users lead to overly conservative assumptions, in which almost every possible driving situation would lead to a collision risk. Such systems would keep generating collision alerts, and would be eventually disregarded or disabled by users.

We thus have to rely not only on the current state of the system, but also on the forecasted user inputs, together with the dynamical model to predict a realistic subset of the paths that can be taken by the user. This prediction can be achieved through the unique sensing capabilities of wearables, which can monitor several body cues, together with the vision of the user (through smart glass cameras). For example, this can be used to check if a user has seen a motorbike passing them on the left (by monitoring the mirrors, using machine vision), and detect if the same user is about to steer left, through body cues such as the flexion of the wrist (detected by the smart watch IMU), motion of the hands (detected by the smart glass camera) or slight turning of the head to the left (detected by the smart glass IMU). All this information would be used in real time to classify the risk of collision and only warn users if a collision is actually imminent.

Learning how to infer the future trajectories of users will be extremely important to ensure that users adopt the technology and more importantly do not disregard the collision warnings and conflict resolution advice generated by the system.

### D. Conflict resolution and actuation

In addition to detecting impending collisions, the system must offer resolution advice to all users in real-time. This type of advice should be easy to compute, take into account the dynamical systems corresponding to different users, and be displayed in an intuitive manner. Human factors have to be taken into account for resolution advice, to ensure that the advice is likely to be followed by their respective users.

Smart glasses are ideal for visualization, since the information can be displayed for instance as a hologram trajectory overlaid on the scene detected by the camera, in addition to visual cues indicating the location of other users to track. Smart glasses can also equip all users, improving the likelihood of successful collision avoidance, unlike existing collision warning systems which only inform drivers. The resolution advice must be shared to all users, with reversals if some of the users to not follow the suggested paths (in a similar way to Reversal Resolution Advisories (RRAs) generated by TCASs in airplanes.

### E. Cybersecurity

As in all safety critical systems, it is essential to guarantee that no user can interfere with the system through for example spoofing or denial of service. One of the vulnerabilities of the prototype system we currently develop is its reliance on GPS systems, which can be easily jammed or worse, spoofed. Communication channels are also vulnerable, false information can be generated by malicious users to create false collision risks, which could in turn cause actual collisions.

While no satisfactory solution to all possible cyberattack scenarios exist to date, all vehicular systems will eventually have to address this issue, which will likely be done through a combination of encryption, fast computing, relative positioning using networks of antennas, and sensor fusion[7], by integrating camera data generated by smart glasses or by in-vehicle cameras.

## IV. Conclusion

The emergence of wearables and personal area networks offer unprecedented opportunities for transportation safety. Through Personal Area Networks (PANs) of wearables, the different road users (pedestrians, bicyclists, motorcyclists, light vehicle drivers and public transportation operators) can track their motion, and intents. The information can be sent to a centralized system (for example a cloud-based system) that detects and resolves conflicts in real-time. This resolution information can then be sent to users through their smartphones, smart watches or smart glasses. Since wearables are purchased by all users, the marginal cost of using these networks for traffic safety applications is extremely low, which can offer an alternative or a complement to more expensive active traffic safety systems onboard vehicles.

Beyond traffic safety applications, these networks of wearables can also be used for other smart transportation applications. For example, human in the loop control could be used to increase throughput in future intersections by sending optimal orders to drivers, pedestrians, bicyclists and autonomous vehicles. Such implementations would also be greatly facilitated by PANs and augmented reality.